\documentclass[10pt]{cai}

\graphicspath{{figs/}}

\begin{document}

\begin{center}

  \title{Generative Enriched Sequential Learning (ESL) Approach for Molecular Design via Augmented Domain Knowledge}
  \maketitle

  \thispagestyle{empty}
  \begin{tabular}{c}
    Mohammad Sajjad Ghaemi\upstairs{\affilone,*}, Karl Grantham\upstairs{$\S$}, Isaac Tamblyn\upstairs{\affiltwo,\affilthree}, Yifeng Li\upstairs{$\S$}, Hsu Kiang Ooi\upstairs{\affilone}\\
   {\small \upstairs{\affilone} National Research Council Canada, Toronto, ON, Canada} \\
   {\small \upstairs{\affiltwo} Department of Physics, University of Ottawa, Ottawa, ON, Canada} \\
   {\small \upstairs{\affilthree} Vector Institute for Artificial Intelligence, Toronto, ON, Canada} \\
   {\small \upstairs{$\S$} Department of Computer Science, Brock University, St. Catharines, ON, Canada}\\
  \end{tabular}
  
  \emails{
    \upstairs{*}corresponding authors: MohammadSajjad.Ghaemi@nrc-cnrc.gc.ca and HsuKiang.Ooi@nrc-cnrc.gc.ca 
    }

  
  \emails{
    }
\end{center}

\begin{abstract}
Deploying generative machine learning techniques to generate novel chemical structures based on molecular fingerprint representation has been well established in molecular design. Typically, sequential learning (SL) schemes such as hidden Markov models (HMM) and, more recently, in the sequential deep learning context, recurrent neural network (RNN) and long short-term memory (LSTM) were used extensively as generative models to discover unprecedented molecules. To this end, emission probability between two states of atoms plays a central role without considering specific chemical or physical properties. Lack of supervised domain knowledge can mislead the learning procedure to be relatively biased to the prevalent molecules observed in the training data that are not necessarily of interest. We alleviated this drawback by augmenting the training data with domain knowledge, e.g. quantitative estimates of the drug-likeness score (QEDs). As such, our experiments demonstrated that with this subtle trick called enriched sequential learning (ESL), specific patterns of particular interest can be learnt better, which led to generating de novo molecules with ameliorated QEDs.\\
\end{abstract}

\begin{keywords}{Keywords:} Generative Models, Sequential Learning, RNN, Molecular Design, Drug Discovery, QED
\end{keywords}
\copyrightnotice

\section{Introduction}
\label{intro}
In recent years, generative machine learning models have gained attention in drug discovery, and molecular design \cite{sanchez2018inverse, blaschke2018application}. However, only a few methodologies are explicitly designed for sequential structures among these models. HMM are conventional stochastic models where hidden states are developed based on Markov property such that conditional probability of each state given the previous state is independent of all the remaining states \cite{rabiner1986introduction}. While the probabilistic nature of HMM requires expectation–maximization like algorithms such as Baum-Welch to estimate the hidden states \cite{baum1970maximization, lucke1996stochastic}, stochastic gradient descent based approaches are more preferred in the realm of large-scale data \cite{newton2018recent}. RNN and its variants such as LSTM and Gated Recurrent Unit (GRU) are deep learning techniques to model the sequential data and take advantage of stochastic gradient descent based optimization to carry out the sequential training more effectively \cite{mangal2019lstm}. In this study, we propose augmenting the training data for ESL with particular domain knowledge, i.e. QED, to guide the training, which learns salient patterns that can be applied to generating new molecular structures with improved QEDs. The QEDs integrate important physicochemical properties (molecular weights, lipophilicity, etc.) that screened for the desirability of molecules as a drug candidate \cite{Bickerton2012}. Regardless of molecular representation, our algorithm takes in any character-based encoding of a molecule, predicting every character generated in a series. Given a starting character and a number of characters in a sequence to produce, the model generates molecular structure in string format, character-by-character. Improvement in learning for this model consists of identifying the correct sequence of characters in a string.

\section{Relevant Prior Works}
\label{relevant}
New molecule generation using machine learning has been primarily based on natural language processing (NLP), where the molecular structures are largely represented by simplified molecular-input line-entry system (SMILES) string framework \cite{Weininger1988}. Previously, graph (convolutional) neural networks and neural message passing methods have been explored to generate novel molecules \cite{Duvenaud2015,Yang2019}. Similarly, techniques such as generative autoencoders e.g., VAE \cite{Kingma2014}, have been adopted leveraging either SMILES strings \cite{Romez-Bombarelli2018} or molecular graphs \cite{Simonovsky2018,Jin2019,Jin2020b}. 
LSTM is an established deep machine learning technique that promotes the use of sequential or time-series data. Its unique inherent “memory” allows LSTM to impact current input and output through previous knowledge inputs. A number of recent molecular generative models also rely on RNN, and specifically, the LSTM architecture \cite{Olivecrona2017,Segler2018}. Among various models for generating molecular structure, e.g., LatentGAN, VAE and HMM, character-based RNN approach was ranked one of the top techniques that can generate novel molecules with high QEDs \cite{polykovskiy2020molecular}. Apart from the high generative power of RNN based techniques, the flexible structure of the RNN model allows us to generate any arbitrary size molecules with varying diversity hyper-parameter to control the novelty of the generated molecules making it more desirable compared to VAE or GAN techniques.

Moreover, supervised learning guided by a hybrid quantitative measure of drug properties and quantum chemistry simulation has not been explored thoroughly. Here, we developed a particular type of sequential model called ESL guided by domain knowledge to be trained on the QM9 dataset \cite{Ramakrishnan2014}. Next, quantum chemistry simulations were performed on the generated molecular library to calculate the Universal Force Field (UFF) and Hartree-Fock (HF) energy values to further analyze the correlation between these energy values and QEDs.

\section{Proposed Method}
\label{method}
The novelty of our proposed ESL method can be mainly described as guiding the sequential learning framework via augmenting the training dataset. The augmentation procedure is inspired by the supervised domain knowledge that can be pulled together from a set of already trained models incrementally. To achieve this unique objective, we allow the ESL method to take advantage of a combination of sequential learning models, derived by LSTM technique using the 1-hot encoded SMILES representation, that were already trained on QM9 data. In this molecular generation process, the length of the molecules is fixed at 14 characters in length with varying diversity coefficients. This molecule size matches the average size of molecules of the QM9 training dataset and takes into consideration the trade-off between higher QED scores and the compute time during the generation process.
The initial generation model produced a set of molecules that are used for enhancing ESL training. A subset of those generated molecules was selected based on their high QEDs. By training the next generation of the model with a subset of these high QEDs molecules, the intent is to bias molecule generation towards molecular design with high QEDs. In this regard, enriched datasets were generated based on high QEDs; specifically, molecules with QEDs $\geq 0.6$ are selected from the newly generated datasets and combined with molecules with the same QEDs $\geq 0.6$ criterion from the QM9 dataset. Next, the ESL training is based on this combined dataset consisting of about 3000 molecules which were just sufficient to augment a new dataset for incremental training. Subsequently, we generated a new dataset, and with these new molecules, we performed quantum chemistry calculations to determine the UFF energy and ground state energy using HF method. UFF is a measurement of the energy of a molecule in kcal/mol units. A molecule's UFF value is equal to the sum of the following terms: bond stretching and angular distortions, bond angle bending, dihedral angle torsion, inversion terms, non-bonded interactions consisting of van der Waals terms, and electrostatic terms. HF method is a variational method that provides the wave function of a many-body system assumed to be in the form of a Slater determinant for fermions and of a product wave function for bosons. We used the pySCF open-source library to calculate the ground state energy of molecules via the HF method. 

In order to use HF to calculate the ground state of a molecule, the pySCF open-source library was employed. It contains valuable functions for defining the geometry of a molecule, determining the basis set, determining optimal spin and charge, and then calculating the energy of a molecule based on those parameters. These HF calculations were found to be heavily sensitive to the geometry of a molecule. However, pySCF does not contain a function that can produce a molecular geometry. Thus, we were dependent on Open Babel to produce Cartesian coordinates. Nevertheless, this method found general success once correct spin and charge parameters could be found for each molecule.

We observed that molecules generated by the ESL model with low ground state energy or potential energy value do not correlate to higher QEDs. However, selected molecules with higher QEDs lie within a specific range of energy values. The relationship between the ground state energy and the potential energy can be represented by the distance between any two points on the natural line, which is the absolute value of the numerical difference of their coordinates. Therefore, taking the absolute value of the calculated energy and its numerical differences, we screened for molecules with higher QEDs.

\section{Experimental Design}
The LSTM method with a sparse categorical cross-entropy loss function was used in this study as a deep learning model for sequential learning implementation in TensorFlow \cite{abadi2016tensorflow}. The trained LSTM model was used later for molecular generation by taking a categorical distribution on an initialized instance to predict the next character returned by the LSTM model. The predicted character was passed as the next input to the LSTM model along with the previous hidden state. This iteration was repeated for a given length of the desired molecule. The initial first generation of the SL framework was trained unsupervised based on the QM9 public dataset \cite{Ramakrishnan2014}, which is consists of 134 000 stable small organic molecules that are made up of Carbon (C), Hydrogen (H), Oxygen (O), Nitrogen (N) and Fluorine (F). From this initial SL framework, we generated two datasets of molecules (CHONF) independently, each consisting of 7000 molecules. The standard length of these molecules in the experiment is set to 14 characters long (SMILES format). We have chosen diversity coefficients that range from 1-3 to generate a wide range of variability in the molecular structures with starting atom such as CHONF. Next, to generate an enriched molecular dataset based on high QEDs, we select molecules with QEDs $\geq$ 0.6 from the two sets of newly generated CHONF molecules. We also screen molecules from the QM9 datasets with QEDs $\geq$ 0.6. Combining these 3 datasets, we have a total of 3000 molecules which was just sufficient for subsequent incremental training of the SL framework. ESL generated molecules achieved  higher QEDs as a direct consequence of domain augmentation through a well balanced trade-off across all the trained models.

\section{Results}

The generated molecules from unsupervised SL model have lower QEDs in Figure~\ref{fig:sample_table}. After incremental training with the enriched datasets, the ESL model generated molecules are improved in their QEDs. Examples of these molecules are depicted in Table \ref{fig:sample_table} labelled as second-generation molecules. Initially, we screened 2735 molecules with QEDs $\geq 0.6$ from the QM9 dataset. Then, utilizing the SL model trained on the QM9 dataset, we generated two new datasets containing 7000 molecules. A total of 167 molecules were screened from the first set for QEDs $\geq 0.6$ while 171 molecules were screened from the second dataset.

\label{results}
\vspace*{-.1in}
\begin{figure}[H]
    \centering
    \includegraphics[scale=0.44]{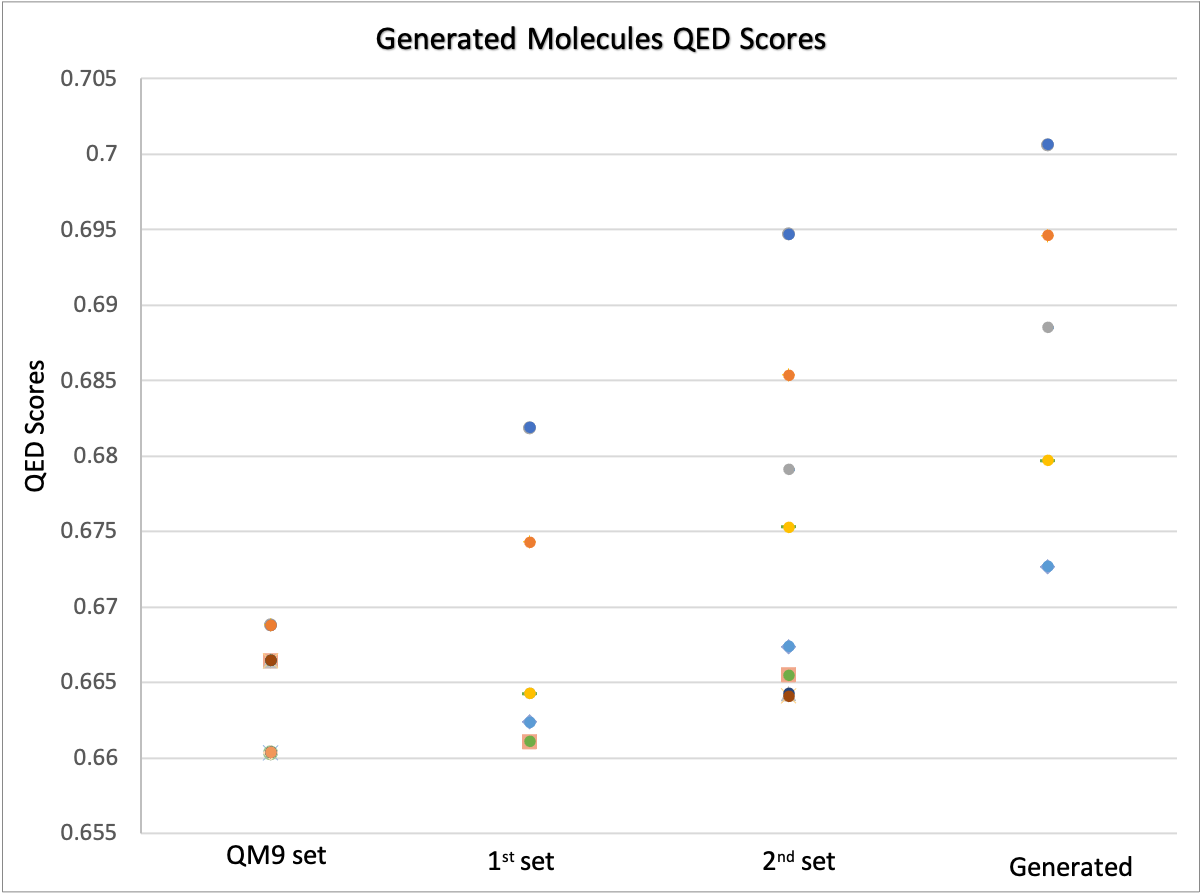} 
    \caption{Screened molecules (each colored datapoint represent an unique molecule) with QEDs $\geq 0.6$ shows improvements in the QEDs properties (left to right). ESL generated molecules with consistently higher QEDs and achieved improved maximum QEDs (i.e. QEDs=0.7006).}
    
    \label{fig:retrain}
\end{figure}
\vspace*{-.1in}

\vspace*{-.1in}
\begin{figure}[H]
    \centering
    \includegraphics[scale=0.4]{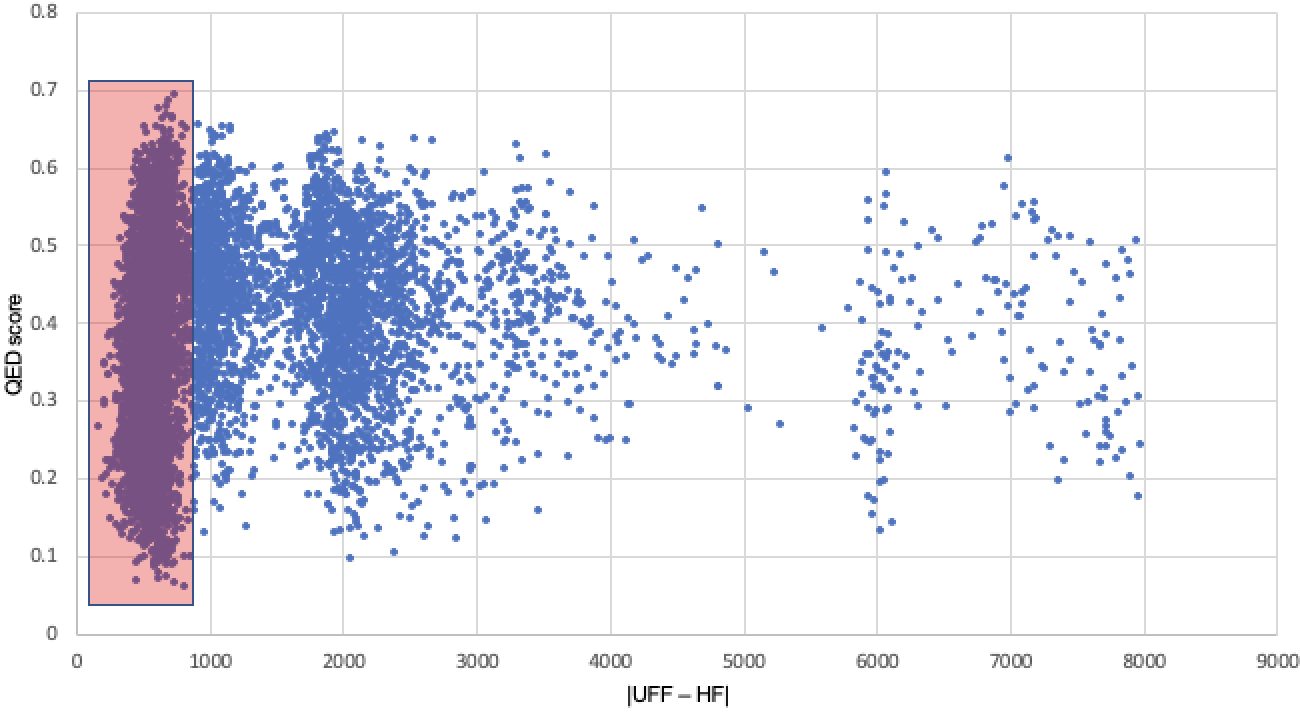} \\
    \caption{Generated molecules with higher maximum QEDs clusters within a specific energy range as highligted in red. The probability of discovering higher QEDs molecules are higher in the dense region of the cluster, by applying the screening criteria of QEDs $\geq 0.6$, we can further generate enriched datasets for incremental training.   }
    \label{fig:uff_hf}
\end{figure}
\vspace*{-.1in}

 Our ESL method resulted in the generation of novel molecules that exhibit molecules with higher maximum QEDs than the unsupervised SL as shown in Figure \ref{fig:retrain} (i.e., Generated). For example, from Figure \ref{fig:retrain}, the maximum QEDs from the QM9 dataset is 0.6688, while the first and second datasets have a maximum QEDs of 0.6819 and 0.6947, respectively. After the incremental training with the screened molecules, the enriched SL model is capable of generating molecule with a maximum QEDs of 0.7006 as shown in Figure \ref{fig:retrain}. Examples of these high QEDs are shown as second generation molecules in Table \ref{fig:sample_table}.
 
 Taking the absolute value of the calculated energy and its numerical differences, we observe a cluster of molecules with higher QEDs as shown in Figure \ref{fig:uff_hf} highlighted in red. This cluster of molecules can be screened to develop another enriched dataset for subsequent incremental training of the ESL framework.
 
\begin{table}[H]
    \centering
    \includegraphics[scale=0.32]{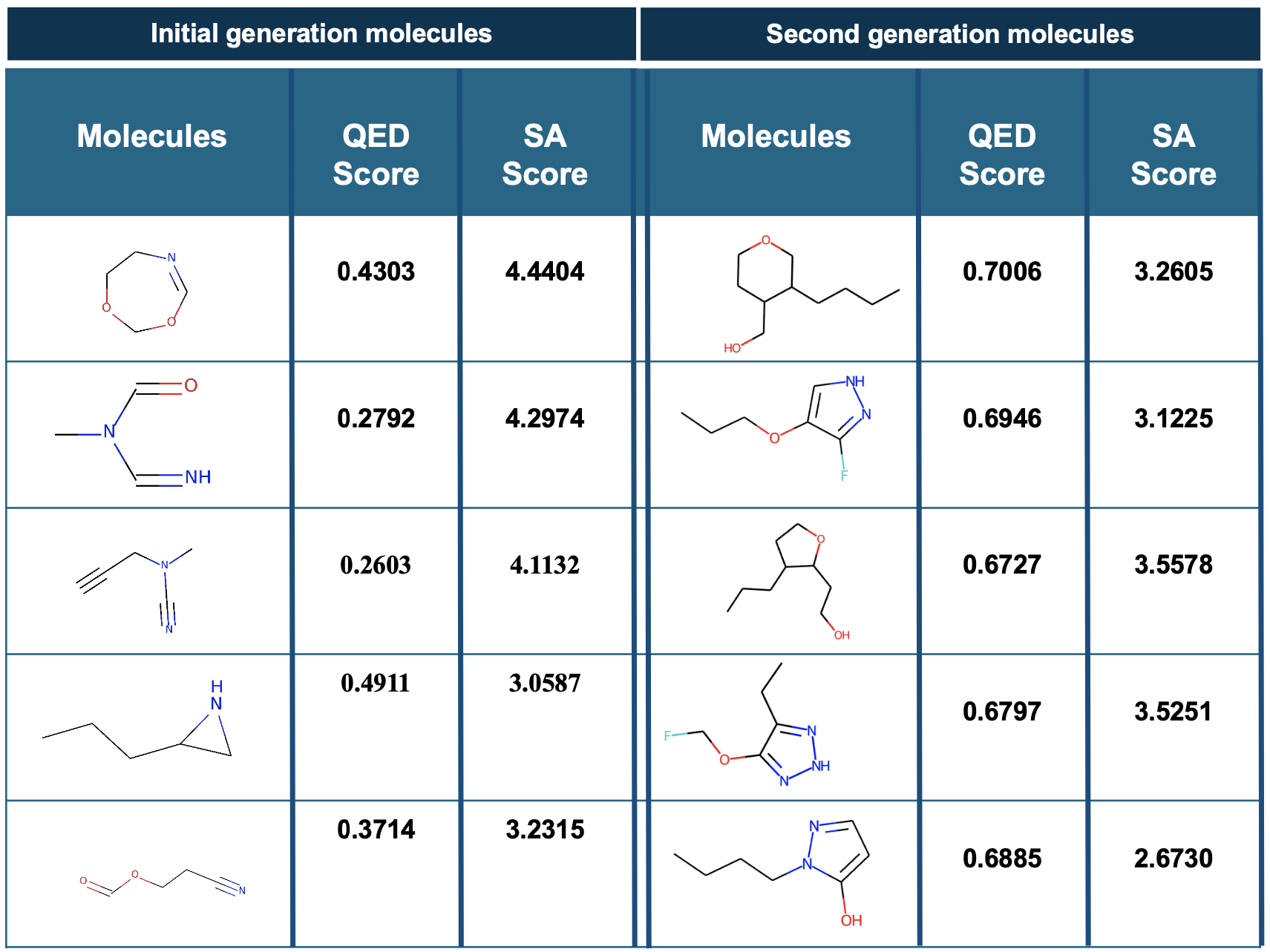}
    \caption{Initial generation (left) of molecules from SL results in poor QEDs. The subsequent incremental training with enriched datasets yields second-generation molecules with higher QEDs (right). The structure of second-generation molecules tend to display a ring structure which suggest the learning converges to a specific design principle.}
    \label{fig:sample_table}
\end{table}
\vspace*{-.2in}

\section{Conclusion and Future Work} 
This approach with successive enriched training datasets has shown that the ESL framework can guide the model towards generating molecules with better properties. Furthermore, while this experiment uses a single objective function (i.e. QEDs), we can adopt multi-objective functions combined with Synthetic Accessibility scores. 

Subsequently, through quantum chemistry simulation, we have shown that additional criteria for screening best-in-class molecules can be used to further enrich datasets for successive training. The HF method leads to future work where the Post-Hartree-Fock method, such as quantum phase estimation (QPE) and variational quantum eigensolver (VQE), can be explored once we have a practical quantum computer to generate similar enriched datasets.

Our contribution highlights a hybrid approach that incorporates a machine learning framework to quantum chemistry simulation that can take advantage of near-term Noisy Intermediate Scale Quantum (NISQ) technology. We are of the opinion that machine learning and quantum computing can be mutually beneficial. Once the quantum computing hardware and software are scalable, we can explore a hybrid quantum machine learning framework leveraging enriched datasets.

\section*{Acknowledgements} This project is supported by the National Research Council Canada under the AI for Design Challenge Program: A1-016449. Karl Grantham is also supported by Vector Scholarship in Artificial Intelligence, provided through the Vector Institute.



\thispagestyle{plain}
\printbibliography[heading=subbibintoc]

\end{document}